\documentclass[draft]{mn2e} 
\title[The 2006 outburst of RS~Oph]
      {Infrared observations of the 2006 outburst of the recurrent nova
      RS~Ophiuchi: the early phase}
\author[A. Evans et al.]
      {A. Evans$^{1}$,  T. Kerr$^{2}$, Bin Yang$^3$,Y. Matsuoka$^4$, Y.
      Tsuzuki$^5$, M. F. Bode$^{6}$, \newauthor  
      S. P. S. Eyres$^{7}$, T. R. Geballe$^{8}$, C. E. Woodward$^{9}$, R. D. Gehrz$^{9}$,
      D. K. Lynch$^{10}$,  \newauthor 
      R. J. Rudy$^{10}$, R. W. Russell$^{10}$, T. J. O'Brien$^{11}$, S. G. Starrfield$^{12}$,  %
      R. J. Davis$^{11}$, \newauthor  
      Jan-Uwe Ness$^{12}$, J. Drake$^{13}$,  J. P. Osborne$^{14}$, K. L. Page$^{14}$, 
      A. Adamson$^2$, \newauthor G. Schwarz$^{15}$,  J. Krautter$^{16}$  
      \mbox{~~} \\ 
      $^{1}$Astrophysics Group, Keele University, Keele, Staffordshire, ST5 5BG, UK \\
      $^{2}$Joint Astronomy Centre, 660 N. A'ohoku Place, University Park,
      Hilo, Hawaii 96720, USA\\
      $^{3}$Institute for Astronomy, University of Hawaii, 2680 Woodlawn Drive,
      Honolulu, HI 96822, USA\\
      $^{4}$Institute of Astronomy, School of Science, The University of Tokyo,
      2-21-1 Osawa, Mitaka, Tokyo 181-0015, Japan\\
      $^{5}$Institute for Cosmic Ray Research, University of Tokyo, 5-1-5,
      Kashiwanoha, Kashiwa, Chiba 277-8582, Japan\\
      $^{6}$Astrophysics Research Institute, Liverpool John Moores University,
      Twelve Quays House, Birkenhead CH41 1LD, UK\\      
      $^{7}$Centre for Astrophysics, University of Central Lancashire,
      Preston, PR1 2HE, UK \\
      $^{8}$Gemini Observatory, 670 N. A'ohoku Place, Hilo, HI\,96720, USA \\
      $^{9}$Department of Astronomy, School of Physics \& Astronomy, 116 Church
      Street S.E., University of Minnesota,\\ \mbox{} \hspace{1mm} Minneapolis, MN 55455, USA \\
      $^{10}$The Aerospace Corporation, Mail Stop M2-266, P.O. Box 92957, Los
      Angeles, CA~90009-2957, USA\\
      $^{11}$Department of Physics \& Astronomy, University of Manchester,
      Manchester, UK \\
      $^{12}$Department of Physics \& Astronomy, Arizona State University,
      Tempe, AZ 85287, USA \\
      $^{13}$Harvard-Smithsonian Center for Astrophysics (CfA), 60 Garden
      Street, Cambridge, MA 02138, USA \\
      $^{14}$Department of Physics and Astronomy, University of Leicester,
      Leicester, LE1 7RH, UK\\      
      $^{15}$Steward Observatory, University of Arizona, 933 North Cherry
      Avenue, Tucson, AZ 85721, USA \\
      $^{16}$Landessternwarte, K\"{o}nigstuhl, D-69117 Heidelberg, Germany }

\date{Revised bersion}

\pagerange{\pageref{firstpage}--\pageref{lastpage}}

\def\LaTeX{L\kern-.36em\raise.3ex\hbox{a}\kern-.15em
    T\kern-.1667em\lower.7ex\hbox{E}\kern-.125emX}

\newcommand{\vunit}{\mbox{\,km\,s$^{-1}$}}
\newcommand{\mic}{\mbox{$\,\mu$m}}
\newcommand{\pion}[2]{{#1}\,{\sc {#2}}}
\newcommand{\fion}[2]{[{#1}\,{\sc {#2}}]}
\newcommand{\ltsimeq}{\raisebox{-0.6ex}{$\,\stackrel 
	{\raisebox{-.2ex}{$\textstyle <$}}{\sim}\,$}} 
\newcommand{\gtsimeq}{\raisebox{-0.6ex}{$\,\stackrel
	{\raisebox{-.2ex}{$\textstyle >$}}{\sim}\,$}}

\newcommand{\rs}{RS~Oph}

\begin{document}
\label{firstpage}
\maketitle

\begin{abstract}
We present infrared spectroscopy of the recurrent nova RS~Ophiuchi,
obtained 11.81, 20.75 and 55.71~days following its 2006 eruption. The
spectra are dominated by hydrogen recombination lines, together with
\pion{He}{i}, \pion{O}{i} and \pion{O}{ii} lines; the electron
temperature of $\sim10^4$K implied by the recombination spectrum suggests
that we are seeing primarily the wind of the red giant, ionized by the
ultraviolet flash when \rs\ erupted. However, strong coronal emission
lines (i.e. emission from fine structure transitions in ions having high
ionization potential) are present in the last spectrum. These imply a
temperature of $930\,000$~K for the coronal gas; this is in line with
x-ray observations of the 2006 eruption. The emission line widths
decrease with time in a way that is consistent with the shock model
for the x-ray emission. 
\end{abstract}


\section{Introduction}

RS~Ophiuchi is a recurrent nova that has undergone nova eruptions
in 1898, 1933, 1958, 1967, 1985, and possibly \cite{1907} 1907. As
in the case of a classical nova, the eruption follows a thermonuclear
runaway on the surface of the white dwarf \cite{tnr}.

The key differences between classical and recurrent novae, in terms of
both system properties and outburst behaviour, are reviewed by Anupama
\shortcite{anupama}. The recurrents are a heterogeneous class of objects
but the \rs\ type is characterized by a semi-detached binary consisting
of a roche-lobe-filling M~giant mass donor (M8III in the case of \rs;
\cite{fekel}) and a massive (close to the Chandrasekhar limit) white
dwarf ({\em classical} novae almost exclusively have cool {\em dwarf}
mass donors).

\begin{table*}
\begin{center}
\caption{Observing log.}
\begin{tabular}{cccccccccc} \hline
2006 UTC Date & Day &\multicolumn{2}{c}{Exp (s)} &  \multicolumn{4}{c}{Resolution} & Comment  \\ 
             &     & $IJ$ & $HK$                     & $I$ & $J$ & $H$ & $K$ &   \\ \hline
Feb 24.64    & 11.81 & 80 & 80 &  350 & 350   & 1000 & 1000 &        \\
Mar 5.58     & 20.75 & 80 & 80 &  350 & 350   & 1500 & 1500 & Cloud  \\
Apr 9.54     & 55.71 & 80 & 80 &  350 & 350   & 1500 & 1500 & Thin cirrus \\ \hline
\end{tabular}
\end{center}
\label{obs}
\end{table*}

\begin{figure*}
\setlength{\unitlength}{1cm}
\begin{center}
\leavevmode
\begin{picture}(5.0,5.0)
\put(0.0,4.0){\includegraphics{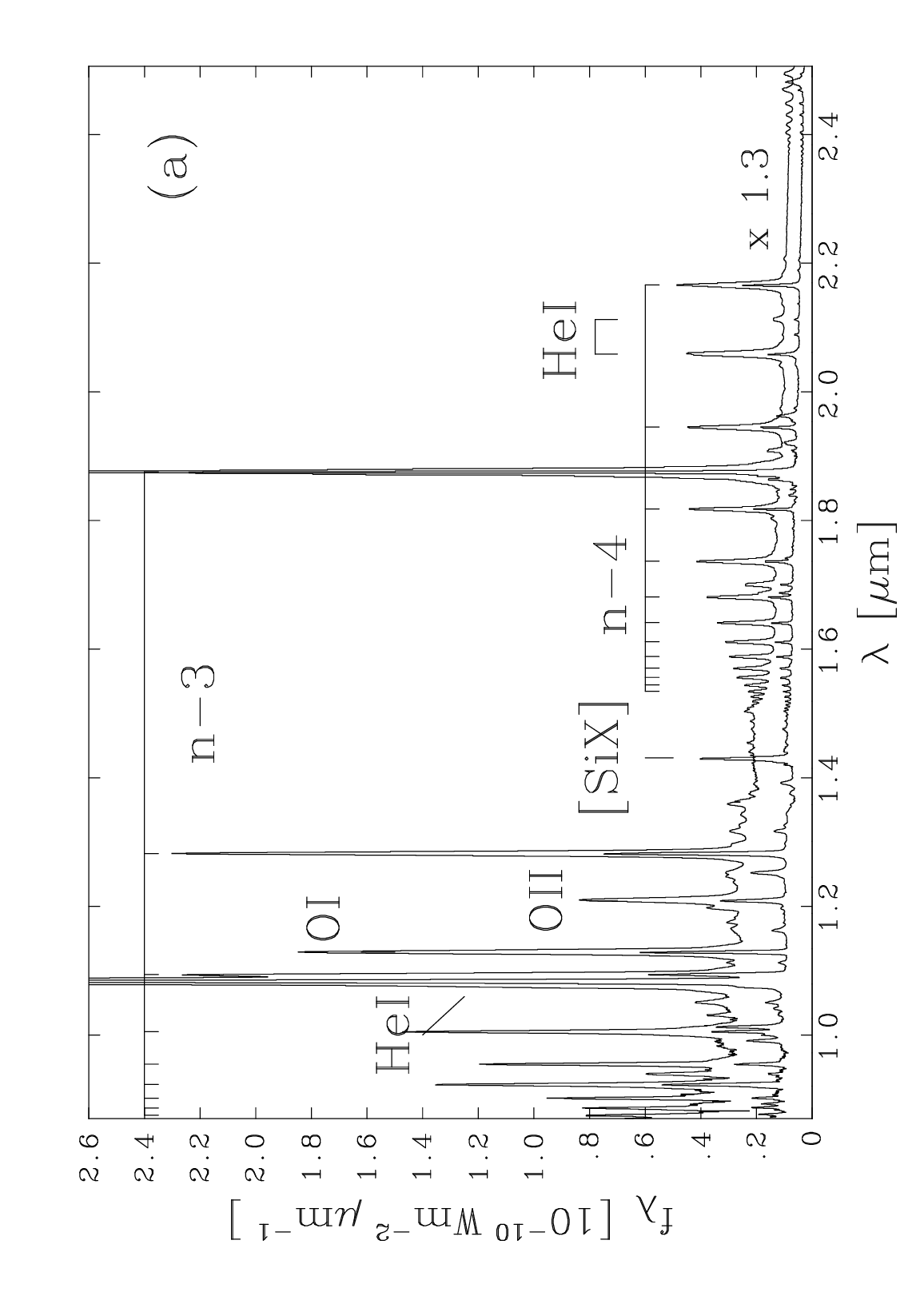}}
\put(0.0,4.0){\includegraphics{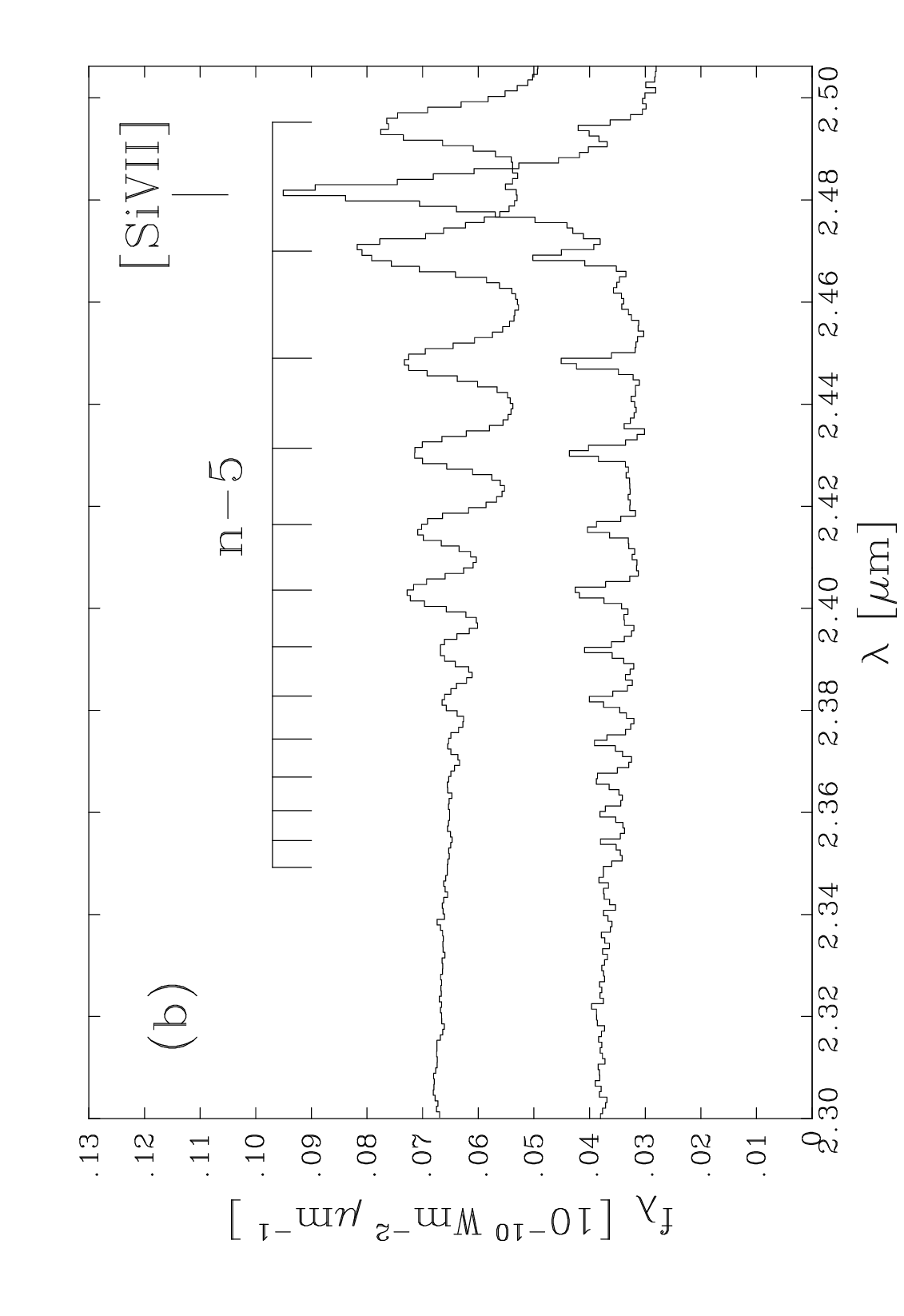}}
\end{picture}
\caption[]{(a) 0.87-2.51\mic\ spectra for Feb 24.64 (upper spectrum, data shifted
upwards by 30\%, for clarity) and April 9.54 (lower spectrum). (b) detail around
2.4\mic\ for Feb 24.64 (upper spectrum) and April 9.54 (lower spectrum).
In both figures the H recombination lines are identified by ``$n-m$''.}
\end{center}
\label{data}
\end{figure*}

The 1985 eruption of \rs\ was the first to have been observed over the
entire electromagnetic spectrum, from the radio to the x-ray (see
Bode 1987). What distinguishes the evolution of the eruption in the
case of \rs\ is the fact that the ejected material runs into the dense
red giant wind, which is shocked \cite{bk}. Observations of the 1985
eruption provided indirect evidence for the shocking of the wind and the
ejecta: \rs\ was a strong and rapidly evolving x-ray \cite{mason} and
radio source \cite{padin}, and there was coronal emission over a wide
range of wavelengths \cite{snijders,evans88,shore}.

Infrared (IR) observations of the 1985 eruption are presented in Evans et al.
\shortcite{evans88}, starting on day~23 of the outburst. These authors found
that the 1--2.5\mic\ spectrum was dominated by hydrogen recombination
lines, and \pion{He}{i}$\lambda$1.083\mic. Coronal lines
(\fion{Si}{vi}$\lambda1.965$\mic\ and \fion{Si}{vii}$\lambda2.481$\mic)
were present on day 143 of the eruption. They also tentatively noted the
presence of first overtone CO emission, although this was based on low
resolution circular variable filter data.

\rs\ was discovered in eruption \cite{hirosawa} on 2006 February
12.83, which we take to be day zero for this outburst. The discovery of
the 2006 eruption triggered a multi-wavelength campaign of observations
\cite{bode-iauc,bode-b,das,evans06,eyres06a,eyres06b,ness-a,ness-b,ness-c,charo,obrien-iauc,obrien-nature}
and yielded for the first time direct evidence, from VLBI imaging,
for an expanding shock \cite{obrien-nature}.

In this paper we present IR spectroscopy of \rs, obtained in the first
55~days of the 2006 eruption, covering a much earlier phase
than that observed in 1985.
 
\section{Observations}

The observations were obtained with the UIST instrument on the United
Kingdom Infrared Telescope (UKIRT), on 2006 February 24, March 5 and
April 9. The data were obtained in the $I\!J\!H\!K$ bands, covering
the wavelength range 0.87--2.51\mic.
First order sky subtraction was achieved by nodding along the slit;
HR6493 was used to remove telluric features and for flux calibration.
Wavelength calibration used an argon arc, and is accurate to $\pm0.0005$\mic\
in the $I\!J$ bands, and to $\pm0.0003$\mic\ in the $HK$ bands.

We note that the precipitable water vapour on March 5 was high
($\sim5$~mm, cf. $\sim1.3$~mm for February 24 and $\sim1.8$~mm
for April 9). As the March 5.58 data were taken through
clouds, and the April 9.54 data were taken through thin cirrus, the
telluric cancellation around 1.85\mic\ (and to a lesser extent
the 1.4\mic\ region) is rather poor, especially for March 5.58.
We estimate that the flux calibration for the February and April
observations is accurate to $\pm20$\%, that for the March observation
is accurate only to $\pm60$\%.

The observing log is given in Table~\ref{obs}, in which the UT times
are times of mid-observation. The data for February 24 and April 9 are
shown in Fig.~\ref{data}; the data for March 5 are omitted in view of the
greater uncertainty in flux calibration.

\begin{figure*}
\setlength{\unitlength}{1cm}
\begin{center}
\leavevmode
\begin{picture}(5.0,5.0)
\put(0.0,4.0){\includegraphics{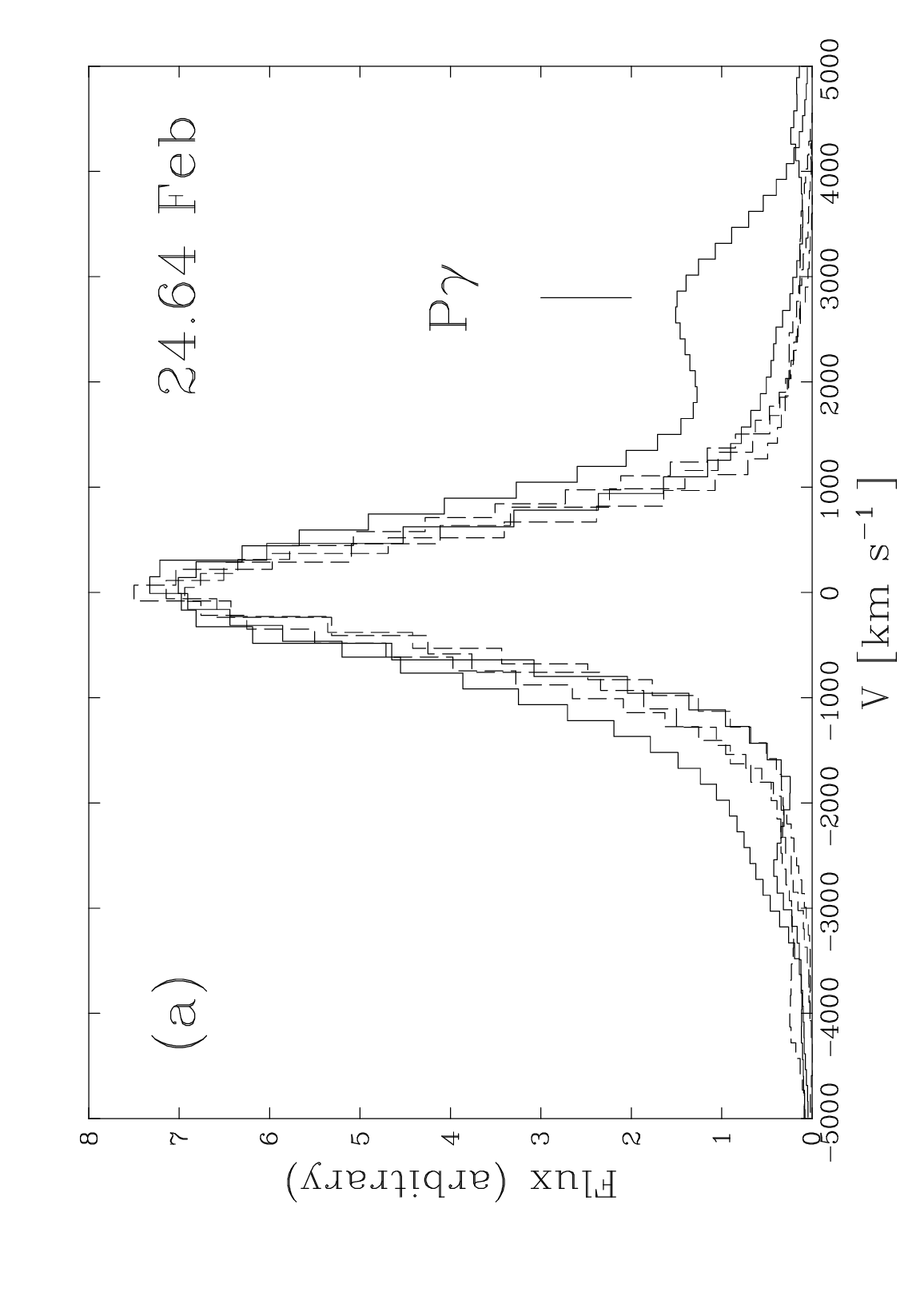}}
\put(0.0,4.0){\includegraphics{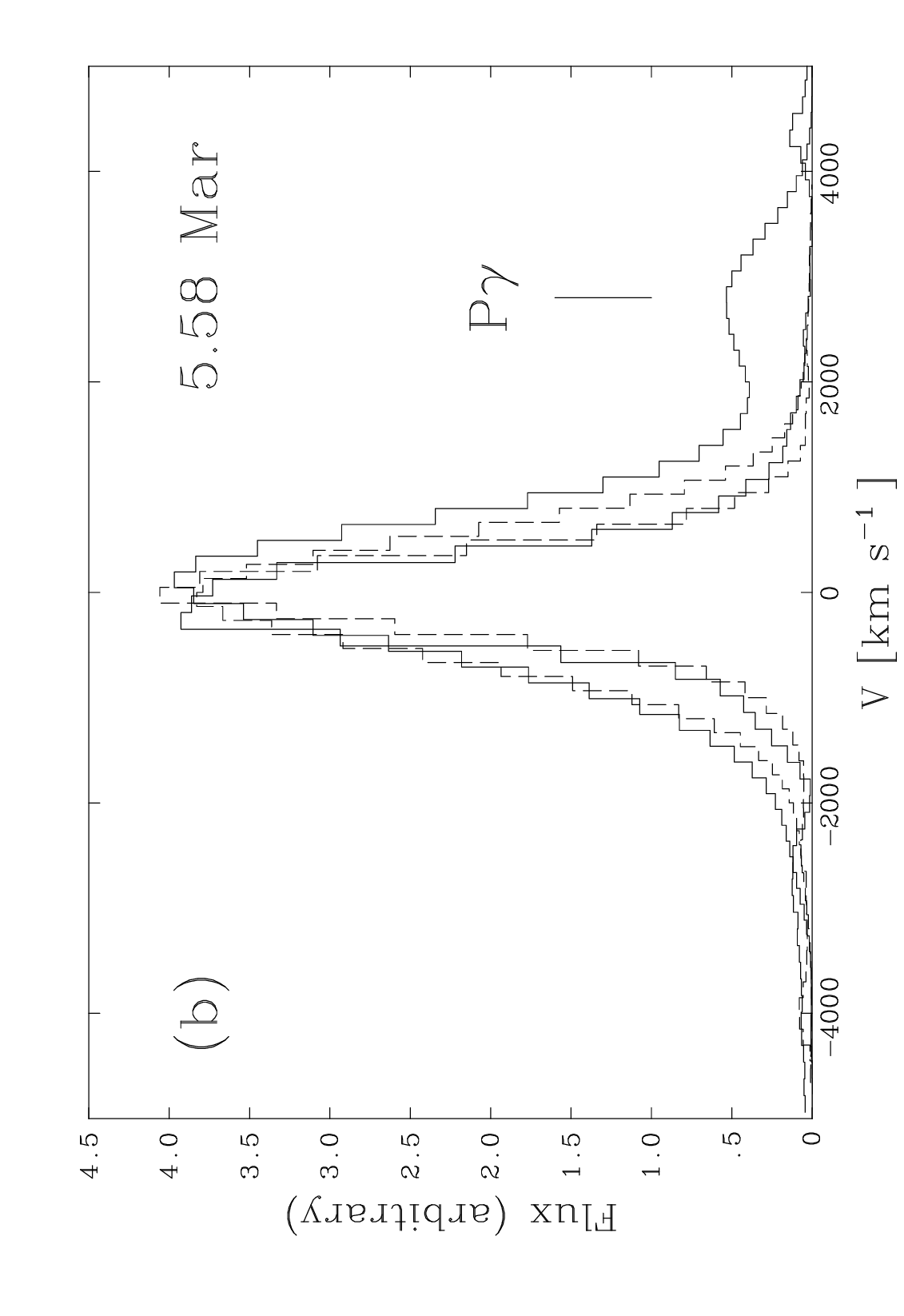}}
\end{picture}
\caption[]{(a) Velocity of selected \pion{H}{i} (P$\alpha$, P$\beta$,
Br$\gamma$; broken lines) and \pion{He}{i} (1.083, 2.058; solid lines)
features on 2006 Feb 24.64 UT;
(b) as (a), but for 2006 Mar 5.58 UT. Note in both cases the wings extend
out to $\sim2500$\vunit. The feature marked P$\gamma$ is the hydrogen
recombination line in the wing of the \pion{He}{i}$\lambda1.083$ line.}
\end{center}
\label{vel}
\end{figure*}

\section{The spectra}

On all dates the spectra are dominated by hydrogen recombination lines,
with \pion{He}{i}, \pion{O}{i} and \pion{O}{ii} also
present (see Table~\ref{nebular}). In particular the higher members of
the hydrogen Pfund ($n\rightarrow5$) series are clearly resolved, and
there is no evidence for the presence of first overtone CO in either
emission or absorption. Thus the identification of CO in the 1985 eruption,
reported by Evans et al. \shortcite{evans88} remains problematic.
By April 9 (day~55) we also see strong emission in silicon
(\fion{Si}{vi}, \fion{Si}{vii}, \fion{Si}{x}) and sulphur (\fion{S}{viii},
\fion{S}{ix}) lines (see Table~\ref{coronal}); the first two of these were
also reported in the 1985 eruption \cite{evans88}.

\begin{figure*}
\setlength{\unitlength}{1cm}
\begin{center}
\leavevmode
\begin{picture}(5.0,5)
\put(0.0,4.0){\includegraphics{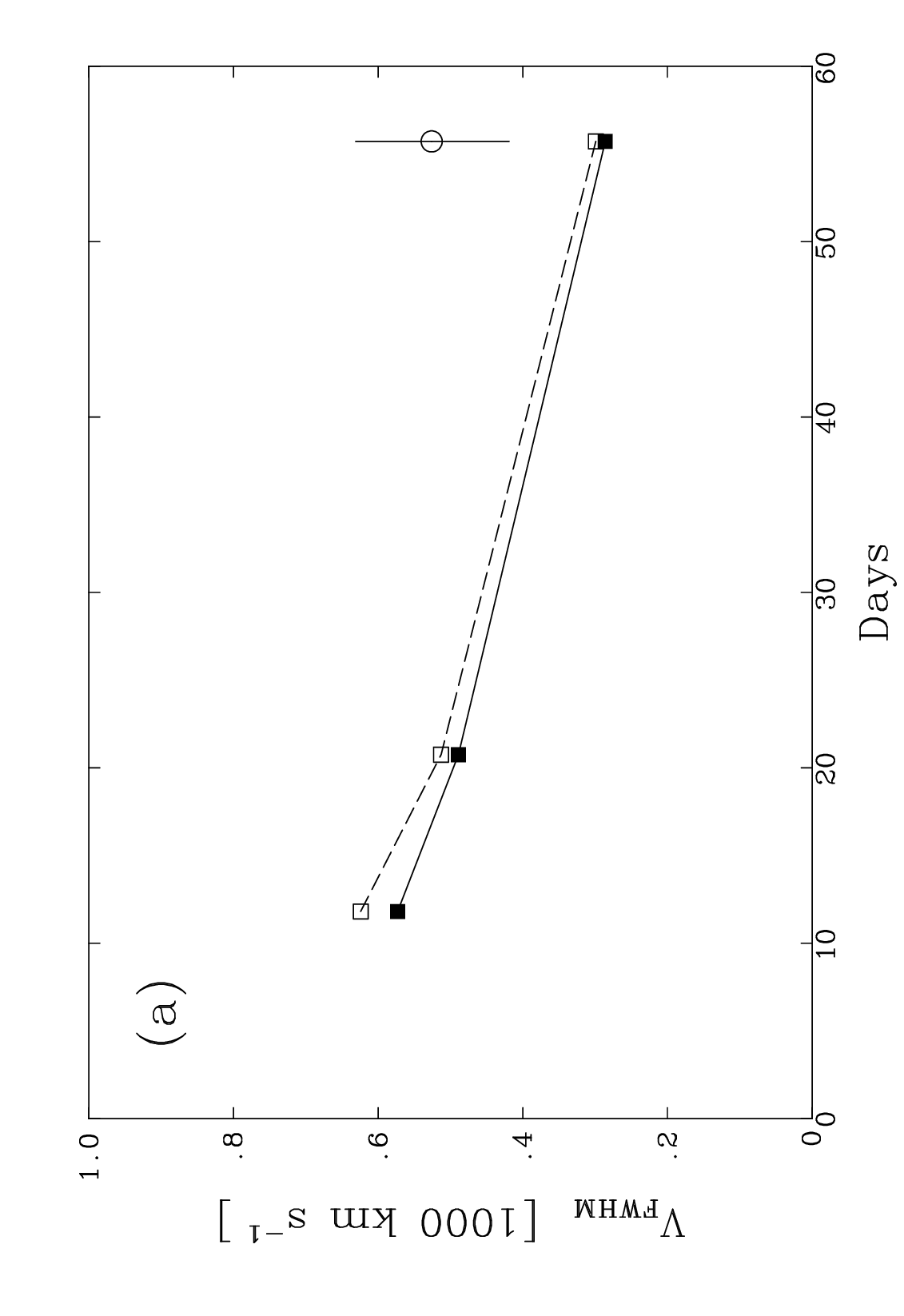}}
\put(0.0,4.0){\includegraphics{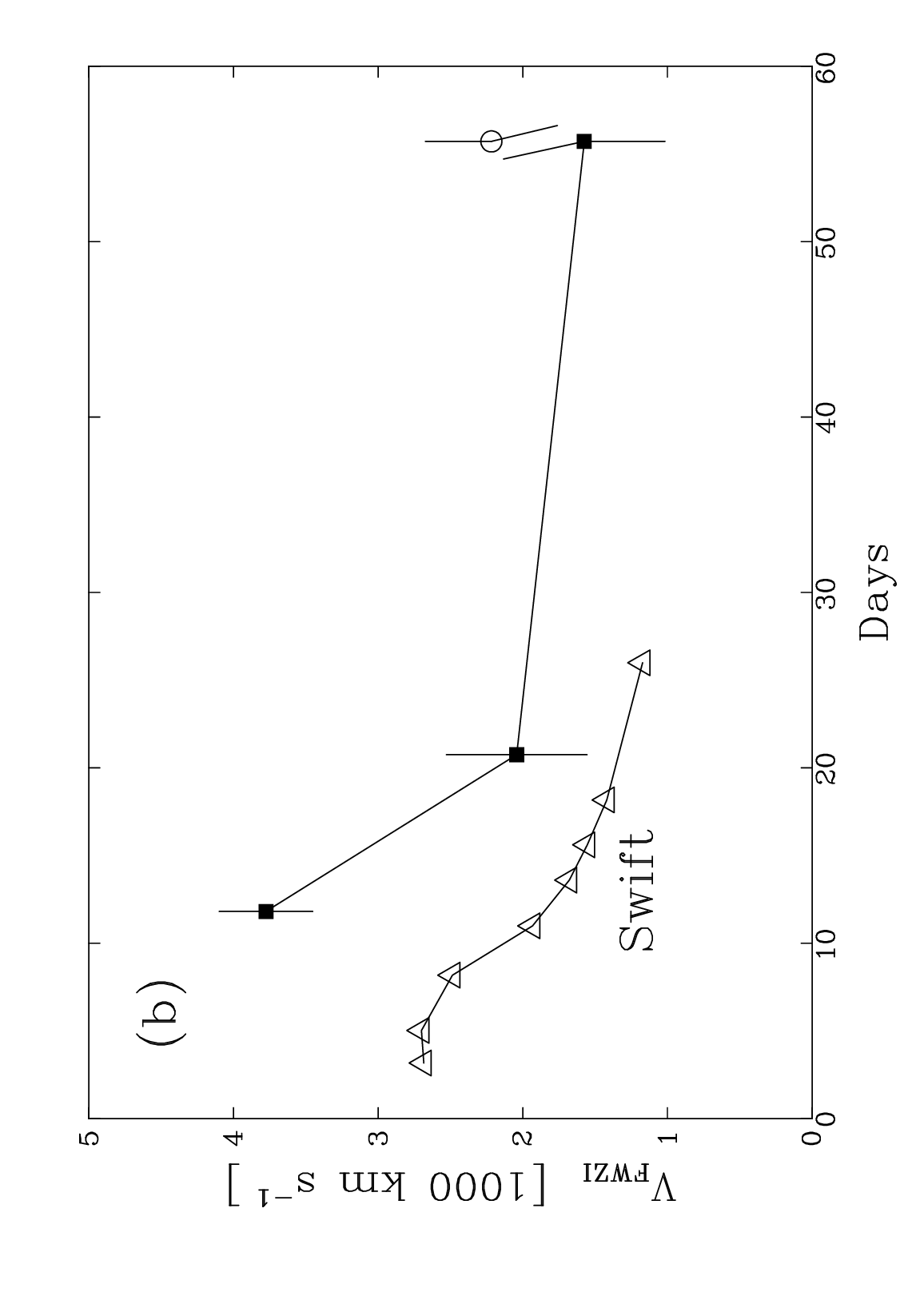}}
\end{picture}
\caption[]{(a) Time-dependence of mean velocity,
as determined from deconvolved FWHM of emission lines. Filled squares and
continuous line, \pion{H}{i} recombination lines only, open squares and
broken line, all nebular lines; dispersion is typically 180\vunit\ (day 11.81),
130\vunit\ (day 20.75), 140\vunit\ (day 55.71). Open circle is mean value
for coronal lines in Table~\ref{coronal}; error bar is dispersion.
(b) As (a), but for mean velocity as determined from FWZI; open circle is
mean FWZI for coronal lines. Error bars are dispersions. Open triangles
show decline of shock velocity as deduced from Swift observations \cite{bode-b}.}
\end{center}
\label{vel_fwzi}
\end{figure*}

We have measured the full-width at half-maximum (FWHM) and full-width
at zero intensity (FWZI) of several emission lines for each of the three
dates (see Figs~\ref{vel},\ref{vel_fwzi}). While the FWHM of the emission
lines indicates an expansion velocity $\sim500-600$\vunit, the emission
line wings extend to $\gtsimeq2500$\vunit. Swift observations \cite{bode-b}
indicate that shock velocities $\sim3000$\vunit\ are present, comparable
with the IR line FWZI. After deconvolving the instrumental linewidth, we
have converted the FWHM to an expansion velocity (cf.
Tables~\ref{nebular},\ref{coronal}) and, from a variety of lines, derived
a mean value for each date.

We find that the mean expansion velocity declines with time, i.e. the
emission lines tend to get narrower as the eruption progresses
(Figs~\ref{vel},~\ref{vel_fwzi}a); furthermore, the velocity
implied by the broad wings also declines (see Fig.~\ref{vel_fwzi}b).
This effect, which arises as the ejected material decelerates as it
ploughs into the giant wind, mirrors the behaviour reported by 
Shore et al. \shortcite{shore} for optical emission lines and by
Snijders \shortcite{snijders} for ultraviolet emission lines during
the 1985 eruption. We note that the velocities determined from the
line wings are comparable with, but somewhat greater than, the shock
velocities deduced from the x-ray emission (see Fig.~\ref{vel_fwzi}).

We also note that the FWHM of the coronal lines in 2006 April (day~55)
is greater than that of the nebular lines. This is clearly seen in
Fig.~\ref{data}b, which includes the \fion{Si}{vii}$\lambda2.483$\mic\ line. 

\begin{table*}
\begin{center}
\caption{Selected nebular lines in \rs; the line fluxes are dereddened
for $E(B-V)=0.73$. Instrumental resolution deconvolved from FWHM values.}
\begin{tabular}{llllcc} \hline
Date & \multicolumn{1}{c}{$\lambda$}  & \multicolumn{2}{c}{Identification} & Dereddened line flux 
& FWHM \\ 
     & \multicolumn{1}{c}{($\!\mic$)} &         &       & ($10^{-14}$~W~m$^{-2}$) & ($\!\vunit$) \\ \hline
Feb 24.64     
     & 1.0833   &  \pion{He}{i}   & $^3$S~ -- $^3$P$^{\rm o}$  & $1131\pm7$   & 949 \\
     & 1.12895  &  \pion{O}{i}    & $^3$P~ -- $^3$D$^{\rm o}$  & $195\pm6.2$  & 720 \\ 
     & 1.2084   &  \pion{O}{ii}   & $^2$D$^{\rm o}$ -- $^2$D~  & $74.0\pm2.0$ & 918 \\
     & 1.7002   &  \pion{He}{i}   & $^3$P$^{\rm o}$ -- $^3$D~  & $9.4\pm0.5$  & 611 \\
     & 2.05869  &  \pion{He}{i}   & $^1$S~ -- $^1$P$^{\rm o}$  & $33.7\pm2.3$ & 659\\    
     &  &  &  &&\\
Mar 5.58  
     & 1.0833   &  \pion{He}{i}   & $^3$S~ -- $^3$P$^{\rm o}$  & $649\pm3.1$  & 710 \\
     & 1.12895  &  \pion{O}{i}    & $^3$P~ -- $^3$D$^{\rm o}$  & $71.8\pm2.6$ & 620 \\ 
     & 1.2084   &  \pion{O}{ii}   & $^2$D$^{\rm o}$ -- $^2$D~  & $30.3\pm3.0$ & 755 \\ 
     & 2.05869  &  \pion{He}{i}   & $^1$S~ -- $^1$P$^{\rm o}$  & $13.47\pm0.96$  & 488\\    
     &  &  &  &&\\
Apr 9.54 
     & 1.0833   &  \pion{He}{i}   & $^3$S~ -- $^3$P$^{\rm o}$  & $405.8\pm2.8$ & 457\\
     & 1.12895  &  \pion{O}{i}    & $^3$P~ -- $^3$D$^{\rm o}$  & $41.9\pm2.5$ & 266\\ 
     & 1.2084   &  \pion{O}{ii}   & $^2$D$^{\rm o}$ -- $^2$D~  & $21.21\pm0.33$ & 420\\
     & 2.05869  &  \pion{He}{i}   & $^1$S~ -- $^1$P$^{\rm o}$  & $5.8\pm0.2$  & 318\\   \hline 
\end{tabular}
\label{nebular}
\end{center}
\end{table*}

\begin{table*}
\begin{center}
\caption{Coronal lines in \rs\ on 2006 April 9; the line fluxes are
dereddened for $E(B-V)=0.73$. Instrumental resolution deconvolved
from FWHM values.}
\begin{tabular}{cllcc} \hline
$\lambda$  & \multicolumn{2}{c}{Identification} & Dereddened line flux & FWHM \\ 
 ($\!\mic$)      &         &       & ($10^{-14}$~W~m$^{-2}$) & ($\!\vunit$) \\ \hline
0.991      & 0.911 \fion{S}{viii} & $^2$P$^{\rm o}$ -- $^2$P$^{\rm o}$    & $14.21\pm0.56$ & 531 \\
1.252      & 1.252 \fion{S}{ix}   & $^3$P -- $^3$P     & $14.97\pm0.39$ & 646 \\
1.431      & 1.430 \fion{Si}{x}   & $^2$P$^{\rm o}$ - $^2$P$^{\rm o}$    & $27.82\pm0.64$ &  355 \\
           & 1.653 \fion{Si}{x}   & $^4$P~ - $^4$P~           & $<0.15$        &  --- \\
           & 1.936 \fion{Si}{xi}  & $^3$P$^{\rm o}$ - $^3$P$^{\rm o}$   & $<0.08$        &  --- \\   
1.962      & 1.965 \fion{Si}{vi}  & $^2$P$^{\rm o}$ - $^2$P$^{\rm o}$   & $5.96\pm2.77$  &  550 \\
2.481      & 2.483 \fion{Si}{vii} & $^3$P~ - $^3$P~           & $6.42\pm0.13$  &  554 \\ \hline
\end{tabular}
\label{coronal}
\end{center}
\end{table*}

\section{Discussion}

The spectra have been dereddened for $E(B-V)=0.73$ \cite{snijders} and the
dereddened fluxes are reported in Table~\ref{nebular} for the He and O
lines and in Table~\ref{coronal} for the coronal lines.

\subsection{The hydrogen recombination lines}

\begin{figure}
\setlength{\unitlength}{1cm}
\begin{center}
\leavevmode
\begin{picture}(5.0,4.5)
\put(0.0,4.0){\includegraphics{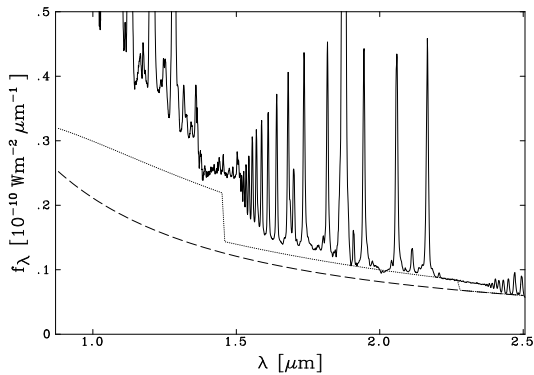}}
\end{picture}
\caption[]{Spectrum for 2006 Feb 24 with nebular continuum at $10^4$~K (dotted
line) and $10^6$~K (broken line).}
\label{cont}
\end{center}
\end{figure}

Assuming that the continuum that is clearly visible in Fig.~\ref{data} is
optically thin free-free and free-bound emission, we estimate the electron
temperature to be $\sim10^4$~K for all three of our observations, but
note that the flux calibration for the March observation is
not reliable as the data were taken through cloud. This temperature is
constrained primarily by the magnitude of the Brackett and Pfund
discontinuities at 1.45\mic\ and 2.28\mic\ respectively (Fig.~\ref{cont}).
The electron temperature derived from the optically thin emission
is considerably less than that implied by the presence of IR coronal lines
in the spectra (see below), or inferred from radio \cite{obrien-nature}
and x-ray \cite{bode-b} observations. There remains an excess at wavelengths
$\ltsimeq1.5$\mic, some (but not all) of which may be due to a contribution
from the shocked gas. 

Using flux ratios for the hydrogen recombination lines, and assuming
Case~B \cite{ferland}, we find that
the electron density for day~55.71 is $\sim10^7$~cm$^{-3}$.
Assuming the mass-loss value given by O'Brien et al. \shortcite{obrien-a},
wind velocity 20\vunit\ and shock velocity $\sim2000$\vunit\ (cf. Fig.~\ref{vel_fwzi}),
the corresponding wind column, integrated from the base of the unshocked wind
to infinity, is $\sim2.0\times10^{21}$~cm$^{-2}$, in good agreement with that
obtained from the x-ray data (e.g. Fig.~3 of Bode et al. \shortcite{bode-b}).

\subsection{The coronal lines}

We can use the dereddened fluxes of the silicon coronal lines for 2006
April 9 to estimate the temperature in the coronal region. The relative
fluxes for lines in a coronal gas are discussed by Greenhouse et al.
\shortcite{greenhouse} and we follow their analysis here, using collisional
strengths from Osterbrock \shortcite{agn2} and Blaha \shortcite{blaha},
and ionization fractions as a function of temperature from Shull \& van
Steenberg \shortcite{shull}. We find that the temperature of the coronal
gas is $\simeq930\,000$~K ($\sim0.08$~keV). We note that
Ness et al. \shortcite{ness-c} deduced a temperature of a few $\times10^6$~K
from the coronal x-ray lines in a Chandra observation on Jun 4.5;
the temperature for the coronal gas obtained here is broadly consistent with
the x-ray data.

\subsection{Origin of the IR emission}

The deduced electron temperature, $\sim10^4$~K, implies that the hydrogen
IR emission on all three dates is primarily due to emission by the red
giant wind, ionized by the ultraviolet flash when \rs\ erupted. Emission
by the shocked wind must also contribute to the total emission; however
IR observational evidence for this is apparent only on day 55.71 with the
clear development of the S and Si coronal
lines. As the shock propagates into, and eventually breaks out of, the wind
(which is predicted to occur around $t\sim70$~days; see O'Brien, Bode
\& Kahn 1992) we expect that the contribution of the
coronal gas will become dominant, and that of the cooler gas to decline
and eventually disappear.

While there exist at least two regions with greatly differing temperatures
in the environment of \rs\ the determination of abundances is problematic.
However we anticipate that this will change when the shock breaks out of the
giant wind. This next phase will be discussed in a forthcoming paper.

\section{Conclusions}

We have reported the early IR spectroscopy of the 2006 eruption of the
recurrent nova \rs, covering the first 55~days. We find a spectrum dominated
by hydrogen recombination lines arising from a gas at $\sim10^4$~K;
silicon coronal lines prominent on day~55, implying a temperature for
the coronal gas of 930\,000~K.

IR (and other) observations of this remarkable object are continuing and
in subsequent papers we will present contemporaneous observations carried
out with UKIRT and the Spitzer Space Observatory.

\section*{ACKNOWLEDGMENTS}

We thank the UKIRT Director and the various UKIRT observers for supporting
this project.
The United Kingdom Infrared Telescope is operated by the Joint Astronomy
Centre on behalf of the U.K. Particle Physics and Astronomy Research Council
(PPARC).
TRG is supported by the Gemini Observatory, which is operated by the Association
of Universities for Research in Astronomy, Inc., on behalf of the international
Gemini partnership of Argentina, Australia, Brazil, Canada, Chile, the United
Kingdom, and the United States of America.
Use of the UKIRT by YM and YT is supported by National Astronomical Observatory
of Japan.
RDG and CEW are supported in part by the NSF (AST02-05814).
The work of DKL, RJR and RWR is supported by The Aerospace Corporation's
Independent Research and Development Program.
J.-U. N. gratefully acknowledges support provided by NASA through
Chandra Postdoctoral Fellowship grant PF5-60039 awarded by the Chandra
X-ray Center, which is operated by the Smithsonian Astrophysical
Observatory for NASA under contract NAS8-03060.
JPO and KLP acknowledge support from PPARC.
SGS acknowledges partial support from NSF grants to Arizona State University.
Data reduction was carried out using hardware and software provided by PPARC.

\bsp

\label{lastpage}

\end{document}